\documentstyle[11pt,psfig,picinpar]{article}

\makeatletter
\renewcommand{\section}{\@startsection{section}{1}{0in}
	{0.4\baselineskip}{0.1\baselineskip}{\Large\bf}}
\renewcommand{\subsection}{\@startsection{subsection}{2}{0in}
	{0.25\baselineskip}{-\baselineskip}{\large\bf}}
\renewcommand{\subsubsection}{\@startsection{subsubsection}{3}{0in}
	{0.1\baselineskip}{-\baselineskip}{\normalsize\bf}}
\makeatother
\begin{document}

%
\makeatletter\newcommand{\ps@icrc}{
\renewcommand{\@oddhead}{\sl{HE 4.2.12}\hfil}}
\makeatother\thispagestyle{icrc}

\begin{center}
{\LARGE \bf Neutrinos and Muons from Atmospheric Charm}
\end{center}

\begin{center}
{\bf L. Pasquali$^{1}$, M. H. Reno$^{2}$, and I. Sarcevic$^{3}$}\\
{\it $^1$Institut f\"ur Physik, Universit\"at Dortmund, D-44221 Dortmund,
Germany\\
$^{2}$Department of Physics and Astronomy, 
University of Iowa, Iowa City, IA 52242, USA\\
$^{3}$Department of Physics, University of Arizona, Tucson, AZ 85721, USA}
\end{center}

\begin{center}
{\large \bf Abstract\\}
\end{center}
\vspace{-0.5ex}

We evaluate the fluxes of neutrinos and muons from charmed particles
produced by cosmic ray interactions with air nuclei. Using next-to-%
leading order perturbative QCD, we compare our results with calculations
based on a PYTHIA Monte Carlo evaluation of charm production. We investigate
the sensitivity of our results to the choice of parton distribution 
functions and the cosmic ray spectrum. Perturbative
QCD yields larger lepton fluxes than the PYTHIA evaluation. The flux
of muons from charm dominates the muon flux from pion and kaon decays
for energies above $\sim 10^5$ GeV.

%

\vspace{1ex}

%
%
\section{Introduction:}
\label{intro.sec}

Neutrinos and muons are produced in the atmosphere from primary cosmic
ray interactions with air nuclei.
The range of energies of the neutrinos and muons determines their
source: at energies below 10-100 TeV, muons
come from pion and kaon decays. Muon neutrinos
and electron neutrinos come from those same decays as well as
from muon decay, depending on the energy. These lepton fluxes are
termed `conventional', in contrast to muons and neutrinos that come
from the decays of charmed particles, denoted `prompt'.

Current interest has been focused on the observations of atmospheric
muon neutrinos and electron neutrinos by the Super-Kamiokande
experiment (Y. Fukuda et al., 1998). Interpretations of these
data rely on the conventional neutrino flux calculations. One interpretation
of the measured deficit of muon neutrinos is that muon neutrinos
are massive and oscillate into tau neutrinos, which are not detected
in the Super-K experiment. A test of this hypothesis would be evidence of tau
neutrino appearance through detection of $\nu_\tau\rightarrow \tau$.
A background flux of tau neutrinos comes from tau neutrinos produced
directly in the atmosphere by charm particle decays, namely, $D_s\rightarrow
\tau\nu_\tau$, followed by the decay of the tau itself. 

Charmed particle contributions to the high energy muon
flux may be the explanation of underground measurements which
appear to be larger than conventional flux calculations predict,
as summarized by Bugaev et al. (Bugaev et al., 1998). The deviation appears
in the TeV energy region for muons.

We present here our results for muon and neutrino fluxes calculated using
next-to-leading order (NLO)
perturbative QCD. We focus on the muon fluxes above 100 GeV, where
we examine the energy at which the crossover from conventional dominated
to prompt dominated flux occurs. In addition, we present our evaluation
of the tau neutrino flux from $D_s$ decays in the atmosphere 
for $\nu_\tau$ energies above 20 GeV and the
corresponding event rates.

\section{Calculational Method:}
\label{format.sec}

The calculation of the prompt lepton fluxes relies on the semi-analytic
method using approximate cascade equations.
Details of the method can be found in Lipari (1993) and in
Pasquali, Reno and Sarcevic (1999). The lepton fluxes depend on the
incident cosmic ray flux. Initially, we assume that the incident cosmic
ray flux is comprised of protons, and that at the top of the atmosphere,
the proton flux $\phi_p(E)=1.7\ (E/{\rm GeV})^{-2.7}$ for
$E<5\cdot 10^6$ GeV and scales as 174 ($E$/GeV)$^{-3}$ for $E\geq 5\cdot
10^6$ GeV,
in units of (GeV cm$^2$ s sr)$^{-1}$. 
This is the same incident flux used by Thunman, Gondolo and
Ingelman (1996) in their evaluation of the prompt lepton fluxes using
the Monte Carlo program PYTHIA.

Of particular importance for charm
production is the $Z$-moment, 
\begin{equation}
Z_{pj}=2f_j\int_0^1 {dx_E\over x_E} {\phi_p(E/x_E)\over \phi_p(E)}
{1\over \sigma_{pA}(E)}{d\sigma_{pA\rightarrow c\bar{c}}(E/x_E)\over dx_E},
\end{equation}
where $x_E=E/E_p$, the energy of the outgoing charmed hadron divided by
the incident nucleon energy, and $f_j$ is the fraction of charmed particles
which emerge as hadron $j$, as outlined in Pasquali, Reno and Sarcevic (1999).

To evaluate the differential cross section for charm production, we chose
a charmed quark mass $m_c=1.3$ GeV as it yields the best consistency with
the experimental data summarized by Frixione, Mangano and
Nason (1998). Our default
parton distribution functions are CTEQ3 (Lai et al., 1995), however, for
comparison, we also used the MRSD$_-$ set (Martin, Stirling \& Roberts,
1993). Our factorization scale ($M$) and renormalization scale ($\mu$)
were chosen as either $m_c$ or $2m_c$, as noted on Fig. 1. To
include NLO corrections, we have parameterized in energy and $x_E$ the
ratio of the NLO to leading order distribution $d\sigma/dx_E$.

\section{Muon Flux:}
\label{session.sec}

\begin{figure}\centerline{
\psfig{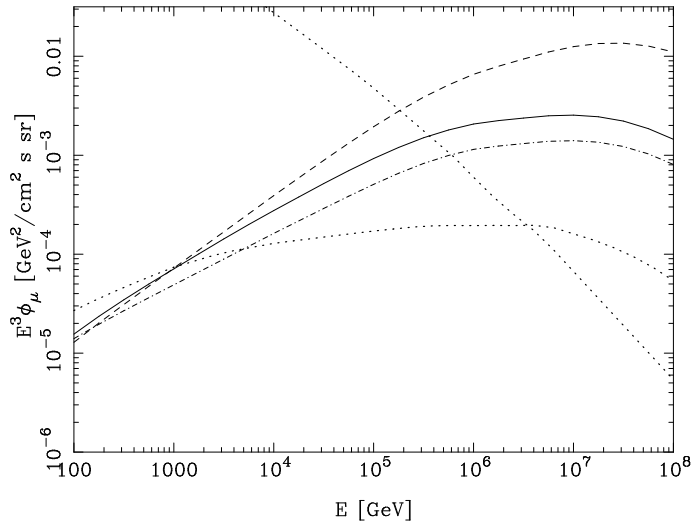}}
\caption{The vertical
prompt atmospheric muon flux scaled by $E^3$ versus muon energy 
for CTEQ3 (solid) and MRSD$_-$ (dashed) with $M = 2\mu = 2m_c$, for
CTEQ3 with $M = \mu = m_c$ 
(dot-dashed) and the Thunman, Gondolo \& Ingelman (1996) 
parameterization of the vertical prompt and conventional muon 
fluxes (dotted).}
\end{figure}
Our results for the vertical prompt atmospheric flux at sea level,
scaled by $E^3$, are shown in Fig. 1 by the solid, dashed and dot-dashed
lines. All fluxes shown are the sum of particle plus antiparticle.
The dotted lines show the Thunman, Gondolo and Ingelman (1996)
vertical prompt and conventional muon fluxes. The prompt flux is 
isotropic below $\sim 10^7$ GeV due to the fact that essentially all
charmed particles decay below that energy. At higher energies, the
turnover in the flux curve indicates that some charmed particles are
not decaying in the region between where they are produced and sea level.

The dashed curve based on the MRSD$_-$ distributions should be considered
as an upper limit, as the parton distribution functions are larger at
small parton $x$ than the results reported by HERA experiments 
(Breitweg et al., 1999,
Adeloff et al., 1997, Derrick et al., 1996). The CTEQ3 distribution
functions are a better fit.
The solid line in Fig. 1 can be parameterized by
\begin{equation}
\log_{10}[E^3\phi_\mu/({\rm GeV}^2/{\rm cm}^2{\rm s\,sr})]
=-5.79+0.345x+0.105x^2-0.0127x^3,
\end{equation}
where $x=\log_{10}(E/{\rm GeV})$. The crossover from conventional to
prompt flux occurs at $E\sim 200-600$ TeV
according to Fig. 1, so perturbative QCD production of
charm in the atmosphere cannot explain the deviation of the muon flux 
from the conventional flux predictions summarized by Bugaev et al.
(1998).

The electron neutrino and muon neutrino fluxes from charm are essentially 
identical to the prompt muon flux shown in Fig. 1. The conventional muon
neutrino
flux is on the order of 1/5 of the conventional muon flux, while the 
conventional electron neutrino flux is $\sim 1/150$ of the conventional
muon flux.

\section{Tau Neutrino Flux:}
\label{additional.sec}

\begin{figure}\centerline{
\psfig{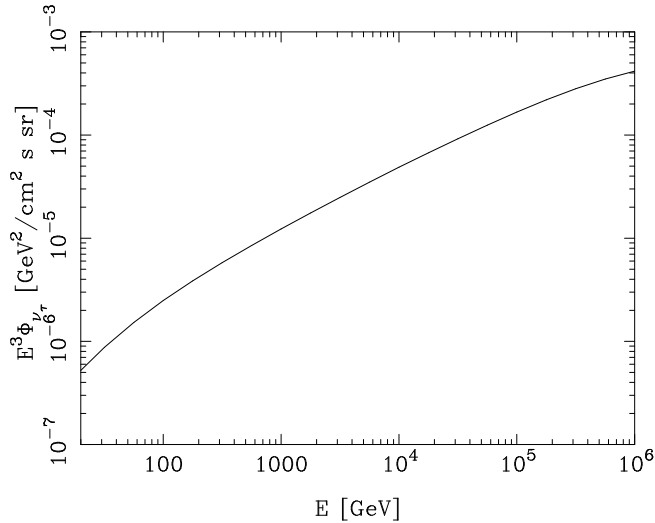}}
\caption{The 
prompt atmospheric tau neutrino flux scaled by $E^3$ versus tau neutrino 
energy.}
\end{figure}

The atmospheric tau neutrino flux, in the absence of neutrino oscillations,
comes primarily from $D_s$ and $\tau$ decays. The branching fraction for
$D_s\rightarrow \tau\nu_\tau$ is taken to be 4.3\% (Gonzalez-Garcia \&
Gomez-Cadenas, 1997), and tau decay channels $\tau\rightarrow \nu_\tau
\mu\nu_\mu,\ \nu_\tau
e\nu_e,\ \nu_\tau \pi,\ \nu_\tau \rho$ and $\nu_\tau a_1$ are included.
A discussion of the calculation is outlined in Pasquali and Reno (1999).
The resulting flux, using the CTEQ3 parton distribution functions and
$M=2\mu=2m_c$ is shown in Fig. 2. The flux is shown for a tau neutrino
energy range from 20 GeV to 10$^6$ GeV. Over this range, the unidimensional
approximation is valid,
the charged particles are relativistic
and geomagnetic corrections are negligible (Agrawal et al., 1996).
The atmospheric tau neutrino flux is isotropic since over the whole
energy range, the decay lengths of the $D_s$ and $\tau$ are short compared
to the height of production.

Event rates for charged current $\nu_\tau+{\rm nucleon}\rightarrow \tau X$ 
conversion are low. In one kilometer-cubed of water equivalent volume,
the number of conversions for tau energies above 20 GeV is 110 events
per year. In a Super-Kamiokande size detector with water volume
of $5\cdot 10^{-5}$ km$^3$, and allowing for a tau range equal to
its time dilated decay length, the event rate per year, 
for the 20 GeV tau energy threshold, is on the order
of 0.55 events. Discussions of higher energy thresholds and the
effects of $\nu_\mu\rightarrow \nu_\tau$ oscillations are found in
Pasquali and Reno (1999).

\section{Discussion:}
\label{title.sec}
We have evaluated the contributions of charmed particle decays to the
atmospheric lepton fluxes. For $\nu_\tau$ fluxes, the $D_s$ decays are
the dominant source. To reduce uncertainties,
the charmed mass was chosen to fit accelerator
data on charm production, however, we have extrapolated our differential
cross sections well beyond the measured regime. In particular, 
the parton distribution functions below parton $x=10^{-5}$ are assumed
to have the form, e.g., $xg(x)\sim x^{-\lambda}$. For $x$
below some critical value $x_c$, this extrapolation should be
invalid due to shadowing effects. 

In an effort to estimate the
effect of shadowing, we have considered $x_c=10^{-6}-10^{-4}$, below
which we have set $\lambda=0.08$. This shadowing affects the
prompt fluxes at $E\sim 10^4-10^6$ GeV, with the lower onset due to 
$x_c=10^{-4}$. It is unlikely that $x_c=10^{-4}$ is realistic in view of
the data from HERA,
including recent measurements at scales above $\sim 1$ GeV 
(Breitweg et al., 1999). At $E=10^8$ GeV,
for $x_c=10^{-4}$ the solid line in Fig. 1 dips to almost
$4\cdot 10^{-4}$, while for $x_c=10^{-6}$,  the flux is
2/3 of the solid curve.
The flattening of the small-$x$ parton distribution
functions doesn't change our qualitative
conclusions about the crossover between
conventional and prompt, however, it does indicate that our fluxes above
$E\sim 10^6$ GeV have significant theoretical uncertainties.

The atmospheric lepton fluxes are directly proportional to the primary cosmic
ray flux at the top of the atmosphere, as well as weakly dependent on 
the primary flux through the $Z$-moments, e.g., as in Eq. (1). Since the 
atmospheric lepton fluxes come from averages of many interactions, the
use of the superposition model, in which a nucleus of mass $A$ is
equivalent to $A$ nucleons, is a good approximation (Engel et al., 1992).
Consequently, what is relevant in the lepton flux calculation is the
number of nucleons per energy-per-nucleon. As indicated in Section 2,
we have used two power laws for the flux
($E^{-2.7}$ and $E^{-3}$), with a critical energy of
$5\cdot 10^6$ GeV for the transition between powers. If we use the
same powers, but move the critical energy to $10^5$ GeV, the new fluxes
begins to deviate from the fluxes in Fig. 1 at $E\sim 10^4$ GeV. The
solid curve drops by a factor of $\sim 1/3$ 
at $E=10^8$ GeV for this choice of
critical energy. The transition from conventional to prompt muons
occurs at a higher energy, on the order of $E\sim 10^6$ GeV.

We thank Chris Quigg for his suggestions for tau neutrino event rates.
Work supported in part by N.S.F. Grant No. PHY-9802403
and D.O.E. Contract No. DE-FG02-95ER40906.

\vspace{1ex}
\begin{center}
{\Large\bf References}
\end{center}
Adeloff, C., et al. 1997, Nucl. Phys. B497, 3\\
Agrawal, V., et al. 1996, Phys. Rev. D 53, 1314\\
Breitweg, J., et al. 1999, Eur. Phys. J. C 7, 609\\    
Bugaev, E. V., et al. 1998, Phys. Rev. D 58, 054001\\
Derrick, M., et al. 1996, Z. Phys. C72, 399\\
Engel, J., et al. 1992, Phys. Rev. D 46, 5013\\
Frixione, S., Mangano, M. L. \& Nason, P. 1998, Heavy Flavours II,
ed. Buras, A. J. \& Lindner, M. (World Scientific)\\
Fukada, Y. (Super-Kamiokande Collaboration) 1998, Phys. Lett. B 433, 9;
436, 33\\
Gonzalez-Garcia, M. C.
\& Gomez-Cadenas, J. J. 1997, Phys. Rev. D 55, 1297\\
Lai, H. et al. 1995, Phys. Rev. D 51, 4763\\
Lipari, P. 1993, Astrop. Phys. 1, 195\\
Martin, A. D., Stirling, W. J. \& Roberts, R. G. 1993, Phys. Lett. B306, 145\\
Pasquali, L., Reno, M. H. \& Sarcevic, I. 1999, Phys. Rev. D 59, 034020\\
Pasquali, L. \& Reno, M. H.  1999, Phys. Rev. D 59, 093003\\
Thunman, M., Gondolo, P. \& Ingelman, G. 1996, Astrop. Phys. 5, 309\\

\end{document}